\documentclass[fleqn,twoside]{article}
\usepackage{espcrc2}

\usepackage{graphicx}
\usepackage[figuresright]{rotating}

\newcommand{\AmS}{{\protect\the\textfont2
  A\kern-.1667em\lower.5ex\hbox{M}\kern-.125emS}}

\title{Status of baryonic $B$ decays}

\author{Hai-Yang Cheng\address[MCSD]{Institute of Physics, Academia
Sinica, Taipei, Taiwan 115, ROC} }

\begin{document}

\def\be{\begin{eqnarray}}
\def\en{\end{eqnarray}}
\def\non{\nonumber}
\def\la{\langle}
\def\ra{\rangle}
\def\nc{N_c^{\rm eff}}
\def\vp{\varepsilon}
\def\drho{\bar\rho}
\def\deta{\bar\eta}
\def\CP{{\it CP}~}
\def\a{{\cal A}}
\def\B{{\cal B}}
\def\c{{\cal C}}
\def\d{{\cal D}}
\def\e{{\cal E}}
\def\p{{\cal P}}
\def\t{{\cal T}}
\def\up{\uparrow}
\def\dw{\downarrow}
\def\vma{{_{V-A}}}
\def\vpa{{_{V+A}}}
\def\smp{{_{S-P}}}
\def\spp{{_{S+P}}}
\def\J{{J/\psi}}
\def\ov{\overline}
\def\Lqcd{{\Lambda_{\rm QCD}}}
\def\pr{{Phys. Rev.}~}
\def\prl{{Phys. Rev. Lett.}~}
\def\pl{{Phys. Lett.}~}
\def\np{{Nucl. Phys.}~}
\def\zp{{Z. Phys.}~}
\def\lsim{ {\ \lower-1.2pt\vbox{\hbox{\rlap{$<$}\lower5pt\vbox{\hbox{$\sim$}
}}}\ } }
\def\gsim{ {\ \lower-1.2pt\vbox{\hbox{\rlap{$>$}\lower5pt\vbox{\hbox{$\sim$}
}}}\ } }

\begin{abstract}
An overview of exclusive two-body and three-body baryonic $B$
decays is given. The threshold enhancement effect in the dibaryon
invariant mass and the angular distributions in the dibaryon rest
frame are stressed and explained. Weak radiative baryonic $B$
decays mediated by the electromagnetic penguin process $b\to
s\gamma$ are discussed. Apart from the first observation of the
radiative penguin decay $B^-\to\Lambda\bar p\gamma$ with the
baryonic final state, the decay $B^-\to \Xi^0\bar\Sigma^-\gamma$
at the level of $6\times 10^{-7}$ may be accessible to $B$
factories in the near future.
 \vspace{1pc}
\end{abstract}

\maketitle

\section{INTRODUCTION}
In this talk we would like to give an overview of the experimental
and theoretical status of exclusive baryonic $B$ decays. The
announcement of the first measurement of the decay modes $p\bar
p\pi^\pm$ and $p\bar p\pi^+\pi^-$ in $B$ decays by ARGUS
\cite{ARGUS} has stimulated extensive theoretical studies during
the period of 1988-1992. However, experimental and theoretical
activities towards baryonic $B$ decays suddenly faded away after
1992. This situation was dramatically changed in the past six
years. Interest in this area was revitalized by many new
measurements at CLEO, Belle and BaBar followed by active
theoretical studies.

\begin{table}[t]
\caption{Branching ratios (in units of $10^{-5}$) of charmful
two-body baryonic $B$ decays. } \label{tab:2bodycharm}
\begin{center}
\begin{tabular}{|  l l |  } \hline
 Decay & Belle \cite{Belle:Lamcp,Belle:Lamcppi,2cbaryon}  \\ \hline
 $\Lambda_c^+\bar p$ &  $2.19^{+0.56}_{-0.49}\pm0.32\pm0.57$~\footnotemark[1] \\
 $\Lambda_c^+\bar\Delta^{--}$ & $<1.9$  \\
 $\Lambda_c^+\bar\Delta_X(1600)^{--}$ & $5.90^{+1.03}_{-0.96}\pm0.55\pm1.53$  \\
 $\Lambda_c^+\bar\Delta_X(2420)^{--}$ & $4.70^{+1.00}_{-0.92}\pm0.43\pm1.22$  \\
 $\Sigma_c^0\bar p$ &  $3.67^{+0.74}_{-0.66}\pm0.36\pm0.95$   \\
 $\Sigma_c(2520)^0\bar p$ &  $<2.7$  \\
 \hline
 $\Xi_c^0(\to\Xi^-\pi^+)\bar\Lambda_c^-$ &
 $4.8^{+1.0}_{-0.9}\pm1.1\pm1.2$  \\
 $\Xi_c^+(\to\Xi^-\pi^+\pi^+)\bar\Lambda_c^-$ &
 $9.3^{+3.7}_{-2.8}\pm1.9\pm2.4$  \\
 \hline
\end{tabular}
\end{center}
\footnotesize{$^1$ The preliminary BaBar result for $\ov
B^0\to\Lambda_c^+\bar p$ is $(2.15\pm0.36\pm0.13\pm0.56)\times
10^{-5}$ \cite{Tetiana}.}
\end{table}

\begin{table}[t]
\caption{Experimental upper limits on the branching ratios of
charmless two-body baryonic $B$ decays. } \label{tab:2body}
\begin{center}
\begin{tabular}{| l l l l | } \hline
  Decay & BaBar \cite{BaBar:pp,BaBar:ppK} & Belle \cite{Belle:2body}
 & CLEO \cite{CLEO:2body} \\ \hline
 $p\bar p$ & $2.7\times 10^{-7}$ & $4.1\times
 10^{-7}$ & $1.4\times 10^{-6}$ \\
 $\Lambda\bar\Lambda$ &  & $6.9\times 10^{-7}$ & $1.2\times 10^{-6}$ \\
 $\Lambda\bar p$ & & $4.9\times
 10^{-7}$ &  $1.5\times 10^{-6}$  \\
 $\Lambda(1520)\bar p$ & $1.5\times 10^{-6}$ & &  \\
 $p\bar\Delta^{--}$ & & & $1.5\times 10^{-4}$  \\
 $\Delta^0\bar p$ & & & $3.8\times 10^{-4}$  \\
 $\Delta^{++}\bar\Delta^{--}$ & & &  $1.1\times 10^{-4}$  \\
 $\Delta^0\bar\Delta^0$ & & & $1.5\times 10^{-3}$ \\
 \hline
\end{tabular}
\end{center}
\end{table}

\begin{table}[thb]
\caption{Branching ratios (in units of $10^{-6}$) of charmless
three-body baryonic $B$ decays. Except for the $p\bar p K^-$ mode,
BaBar results are all preliminary.} \label{tab:3charmless}
\begin{center}
\begin{tabular}{| l l l |  } \hline
 Mode~~ & BaBar \cite{Tetiana,BaBar:ppK}  & Belle \cite{Belle:3charmless,Belle:3charmless1}  \\ \hline
 $p\bar p K^-$ & $6.7\pm0.5\pm0.4$ &
 $5.30^{+0.45}_{-0.39}\pm0.58$ \\
 $p\bar p \ov K^0$ & $2.95\pm0.53\pm0.26$ & $1.20^{+0.32}_{-0.22}\pm0.14$ \\
 $p\bar p K^{*-}$  & $4.94\pm1.66\pm1.00$ &  $10.31^{+3.62+1.34}_{-2.77-1.65}$ \\
 $p\bar p\ov K^{*0}$ & $1.28\pm0.56^{+0.18}_{-0.17}$ & $<7.6$ \\
 $p\bar p\pi^-$ & $1.24\pm0.32\pm0.10$ &  $3.06^{+0.73}_{-0.62}\pm0.37$ \\
 $\Lambda\bar p\pi^+$ & $3.30\pm0.53\pm0.31$ & $3.27^{+0.62}_{-0.51}\pm0.39$ \\
 $\Lambda\bar\Lambda K^-$ & & $2.91^{+0.90}_{-0.70}\pm0.38$ \\
 $\Lambda\bar\Lambda \pi^-$ & & $<2.8$ \\
 $\Lambda \bar pK^+$ & & $<0.82$ \\
 $\Sigma^0\bar p\pi^+$ &  &  $<3.8$ \\
 \hline
\end{tabular}
\end{center}
\end{table}

\section{EXPERIMENTAL STATUS}
\subsection{Two-body decay}

The experimental results for two-body baryonic $\ov B^0$ and $B^-$
decays are summarized in Tables  \ref{tab:2bodycharm} and
\ref{tab:2body} for charmful and charmless decays, respectively.
It is clear that the present limit on charmless ones has been
pushed to the level of $10^{-7}$. In contrast, four of the
charmful 2-body baryonic $B$ decays have been observed in recent
years; among them $\ov B^0\to\Lambda_c^+\bar p$ is the first
observation of the 2-body baryonic $B$ decay mode by Belle
\cite{Belle:Lamcp}. The BaBar's preliminary result for this mode
agrees well with Belle (see \cite{Tetiana} for details). The
decays with two charmed baryons in the final state were measured
by Belle recently \cite{2cbaryon}. Taking the theoretical
estimates (see e.g. Table III of \cite{CT93}), $\B(\Xi_c^0\to
\Xi^-\pi^+)\approx 1.3\%$ and $\B(\Xi_c^+\to \Xi^0\pi^+)\approx
3.9\%$ together with the experimental measurement $\B(\Xi_c^+\to
\Xi^0\pi^+)/\B(\Xi_c^+\to \Xi^-\pi^+\pi^+)=0.55\pm0.16$
\cite{PDG}, it follows from Table \ref{tab:2bodycharm} that
  \be
 \B(B^-\to\Xi_c^0\bar\Lambda_c^-) &\approx& 4.8\times 10^{-3}, \non
 \\
 \B(\ov B^0\to\Xi_c^+\bar\Lambda_c^-)&\approx & 1.2\times 10^{-3}.
 \en
Therefore, the two-body doubly charmed baryonic $B$ decay $B\to
\B_c\bar \B'_c$ has a branching ratio of order $10^{-3}$. Hence,
we have the pattern
 \be \label{eq:2bodypattern}
 \B_c\bar \B'_c~(\sim 10^{-3}) &\gg& \B_c\bar\B~(\sim 10^{-5})\non \\
 &\gg&
 \B_1\bar \B_2~(\lsim 10^{-7})
 \en
for two-body baryonic $B$ decays.

\subsection{Three-body decay}

The measured branching ratios of charmful baryonic decays with one
charmed meson or one charmed baryon or two charmed baryons in the
final state are summarized in Table IV of \cite{Chengreview}. In
general, Belle results are slightly smaller than the CLEO
measurements. The decay $B^-\to J/\psi\Lambda\bar p$ was first
measured by BaBar \cite{BaBar:JLamp} and confirmed by Belle
recently \cite{Belle:JLamp}. In general, $\B(B\to
\B_c\bar\B'_cM)\sim {\cal O}(10^{-3})$ and $\B(B\to \B_c\bar\B
M)\sim {\cal O}(10^{-4})$. The decay $B\to J/\psi\Lambda\bar p$
with the branching ratio of order $10^{-5}$ is suppressed due to
color suppression.

For the charmless case, Belle \cite{Belle:3charmless} has observed
6 different modes while BaBar has tried to catch up, see Table
\ref{tab:3charmless}. The channel $B^-\to p\bar p K^-$ announced
by Belle nearly four years ago \cite{Belle:ppK} is the first
observation of charmless baryonic $B$ decays. Recently Belle has
studied the baryon angular distributions in the baryon-antibaryon
pair rest frame \cite{Belle:3charmless1}, while BaBar has measured
the Dalitz plot asymmetries in the decay $\ov B\to p\bar ph$ with
$h=K^{(*)-,0},\pi^-$. These measurements provide valuable
information on the decay dynamics.

It is naively expected that $p\bar p K^{*-}<p\bar p K^-$ due to
the absence of $a_6$ and $a_8$ penguin terms contributing to the
former and that $p\bar pK^{(*)-}>p\bar p\bar K^{(*)0}$ due to the
absence of external $W$-emission in the latter.  From Table
\ref{tab:3charmless} we see that these expectations are confirmed
except that the Belle measurement of $p\bar pK^{*-}$ seems to be
too large.

There are two common and unique features for three-body $B\to
\B_1\ov \B_2 M$ decays: (i) The baryon-antibaryon invariant mass
spectrum is peaked near the threshold area, and (ii) many
three-body final states have rates larger than their two-body
counterparts; that is, $\Gamma(B\to\B_1\ov
\B_2M)>\Gamma(B\to\B_1\ov \B_2)$. The low-mass enhancement effect
indicates that the $B$ meson is preferred to decay into a
baryon-antibaryon pair with low invariant mass accompanied by a
fast recoil meson. As for the above-mentioned second feature, it
is by now well established experimentally that
 \be
 \B(B^-\to p\bar p K^-) &\gg &  \B(\ov B^0\to p\bar p), \non \\
 \B(\ov B^0\to \Lambda\bar p\pi^-) &\gg&  \B(B^-\to \Lambda\bar p),
 \non \\
 \B(B^-\to\Lambda_c^+\bar p\pi^-) &\gg& \B(\ov B^0\to\Lambda_c^+\bar
 p),  \non \\
 \B(B^-\to\Sigma_c^0\bar p\pi^0) &\gg& \B(B^-\to\Sigma_c^0\bar p).
 \en
This phenomenon can be understood in terms of the threshold
effect, namely, the invariant mass of the dibaryon is preferred to
be close to the threshold. The configuration of the  two-body
decay $B\to\B_1\ov \B_2$ is not favorable since its invariant mass
is $m_B$. In $B\to \B_1\ov\B_2 M$ decays, the effective mass of
the baryon pair is reduced as the emitted meson can carry away
much energies.

An enhancement of the dibaryon invariant mass near threshold has
been observed in several charmless and charmful decays. Threshold
enhancement was first conjectured by Hou and Soni \cite{HS},
motivated by the CLEO measurement of $B\to D^*p\bar n$ and
$D^*p\bar p\pi$ \cite{CLEO:Dpp}. They argued that in order to have
larger baryonic $B$ decays, one has to reduce the energy release
and  at the same time allow for baryonic ingredients to be present
in the final state. This is indeed the near threshold effect
mentioned above. Of course, one has to understand the underlying
origin of the threshold peaking effect. Hence, the smallness of
the two-body baryonic decay $B\to\B_1\ov\B_2$ has to do with its
large energy release.

\section{Theoretical Progress}
\subsection{Two-body decay}

The quark diagrams for two-body baryonic $B$ decays consist of
internal $W$-emission diagram, $b\to d(s)$ penguin transition,
$W$-exchange for the neutral $B$ meson and $W$-annihilation for
the charged $B$. Just as mesonic $B$ decays, $W$-exchange and
$W$-annihilation are expected to be helicity suppressed.
Therefore, the two-body baryonic $B$ decay $B\to\B_1\ov \B_2$
receives the main contributions from the internal $W$-emission
diagram for tree-dominated modes and the penguin diagram for
penguin-dominated processes. It should be stressed that, contrary
to mesonic $B$ decays, internal $W$ emission in baryonic $B$
decays is not necessarily color suppressed. This is because the
baryon wave function is totally antisymmetric in color indices.

Consider the charmful modes $\ov B\to\Xi_c\bar\Lambda_c$ and $\ov
B\to\Lambda_c\bar p$. They have the same CKM angles apart from a
sign difference. Therefore, it is expected that
 \be
 && \B(\ov B^0\to\Lambda_c^+\bar p)=\B(\ov
 B^0\to\Xi_c^+\bar\Lambda_c^-) \non \\
 &&~~\times({\rm dynamical~suppression}).
 \en
where CKM stands for the relevant CKM angles and the dynamical
suppression arises from the larger c.m. momentum in
$\Lambda_c^+\bar p$ than in $\Xi_c\bar\Lambda_c$. Eq.
(\ref{eq:2bodypattern}) implies that the dynamical suppression
effect is of order $10^{-2}$. Likewise,
 \be
 \B(B^-\to\Lambda\bar p)&=& \B(\ov
 B^0\to\Lambda_c^+\bar p)|V_{ub}/V_{cb}|^2 \non \\
 && \times ({\rm
 dynamical~suppression})'  \\
 &\sim& 2\times 10^{-7}({\rm dynamical~suppression})'. \non
 \en
If the dynamical suppression of $\Lambda\bar p$ relative to
$\Lambda_c\bar p$ is similar to that of $\Lambda_c\bar p$ relative
to $\Xi_c\bar\Lambda_c$, the branching ratio of the charmless
two-body baryonic $B$ decays can be even as small as $10^{-9}$. If
this is the case, then it will be hopeless to see any charmless
two-body baryonic $B$ decays.

Earlier calculations based on QCD sum rules \cite{Chernyak} or the
diquark model \cite{Ball} all predict that $\B(B\to
\Xi_c\bar\Lambda_c)\approx \B(\ov B\to\B_c\ov N)$, which is in
strong disagreement with experiment. This implies that some
important dynamical suppression effect for the $\B_c\ov N$
production with respect to $\Xi_c\bar\Lambda_c$ is missing in
previous studies. Recently, this issue was investigated in
\cite{CCT}. Since the energy release is relatively small in
charmful baryonic $B$ decay, the $^3P_0$ model for $q\bar q$
production is more relevant. In the work of \cite{CCT}, the
possibility that the $q\bar q$ pair produced via light meson
exchanges such as $\sigma$ and pions is considered. The $q\bar q$
pair created from soft nonperturbative interactions tends to be
soft. For an energetic proton produced in 2-body $B$ decays, the
momentum fraction carried by its quark is large, $\sim {\cal
O}(1)$, while for an energetic charmed baryon, its momentum is
carried mostly by the charmed quark. As a consequence, the doubly
charmed baryon state such as $\Xi_c\bar\Lambda_c$ has a
configuration more favorable than $\Lambda_c\bar p$. For the
latter, two hard gluons are needed to produce an energetic
antiproton as noticed before: one hard gluon for kicking the
spectator quark of the $B$ meson to make it energetic and the
other for producing the hard $q\bar q$ pair. It is thus expected
that $\Gamma(\ov B\to \B_c\bar N)\ll \Gamma(\ov
B\to\Xi_c\bar\Lambda_c)$ as the former is suppressed by order of
$\alpha_s^4$. This accounts for the dynamical suppression of the
$\Lambda_c\bar p$ production relative to $\Xi_c\bar\Lambda_c$.

\subsection{Three-body decay}
Contrary to the two-body baryonic $B$ decay, the three-body decays
do receive factorizable contributions that fall into two
categories: (i) the transition process with a meson emission, $\la
M|(\bar q_3 q_2)|0\ra\la \B_1\ov \B_2|(\bar q_1b)|B\ra$, and (ii)
the current-induced process governed by the factorizable amplitude
$\la \B_1\ov \B_2|(\bar q_1 q_2)|0\ra \la M|(\bar q_3 b)|B\ra$.
The two-body matrix element $\la \B_1\ov \B_2|(\bar q_1 q_2)|0\ra$
in the latter process can be either related to some measurable
quantities or calculated using the quark model. The
current-induced contribution to three-body baryonic $B$ decays has
been discussed in various publications \cite{CHT01,CHT02,CH03}. On
the contrary, it is difficult to evaluate the three-body matrix
element in the transition process and in this case one can appeal
to the pole model \cite{CYcharmless,CKcharm,CKDmeson}.

Current-induced three-body baryonic $B$ decays such as $\ov
B^0\to\Lambda\bar p\pi^+$ provide an ideal place for understanding
the threshold enhancement effects. It receives the dominant
factorizable contributions from tree  and penguin diagrams with
the amplitude
 \be
&& A(\ov B^0\to\Lambda\bar p\pi^+) = {G_F\over\sqrt{2}}
\la\pi^+|(\bar ub)|\ov B^0\ra  \non \\
&& \times\Big\{ (V_{ub}V^*_{us}a_1
  -V_{tb}V^*_{ts}a_4)\la\Lambda\bar p|(\bar
su)|0\ra \non \\
&& +2a_6V_{tb}V^*_{ts}{(p_\Lambda+p_{\bar p})\over
m_b-m_u}\la\Lambda\bar p|\bar s(1+\gamma_5)u|0\ra\Big\}.
 \en
Based on the pQCD counting rule, the vacuum to $\Lambda\bar p$
form factor has the asymptotic form
 \be
 F(t)\to {a\over t^2}+{b\over t^3}
 \en
in the limit of large $t$. The threshold enhancement effect is
thus closely related to the asymptotic behavior of various form
factors, namely, they fall off fast with the dibaryon invariant
mass. A detailed study in \cite{Tsai} shows that the differential
decay rate for $\Lambda\bar p\pi^+$ should be in the form of a
parabola that opens downward. This is indeed confirmed by
experiment \cite{Belle:3charmless1}: the pion has no preference
for its correlation with the $\Lambda$ or the $\bar p$.

The fragmentation picture advocated in \cite{Rosner} provides some
qualitative descriptions of the correlation in three-body baryonic
$\bar B$ decays. However, some of the predictions based on the
fragmentation mechanism are not borne out by experiment. For
example, the argument in \cite{Rosner} that the $\bar p$ and
$\pi^+$ are neighbors in the fragmentation chain of $\ov
B^0\to\Lambda\bar p\pi^+$ so that the $\pi^+$ is correlated more
strongly to the $\bar p$ than to the $\Lambda$ will lead to an
asymmetric angular distribution which is opposite to what is seen
experimentally.

Apart from the purely transition-induced decays such as $\ov
B^0\to D^{(*)0}p\bar p$, $\ov B^0\to\Sigma_c^{++}\bar p\pi^-$,
most other decays receive both current- and transition-induced
contributions. In the absence of theoretical guidance for the form
factors in the three-body matrix element $\la \B_1\ov \B_2|(\bar
q_1b)|\ov B\ra$, one may consider a phenomenological pole model at
the hadron level as put forward in \cite{CYcharmless}. The meson
pole diagrams are usually related to the vacuum to $\B_1\bar\B_2$
transition form factors and hence responsible for threshold
enhancement, whereas the baryon pole diagrams account for the
correlation of the outgoing meson with the baryon. Indeed, a
detailed analysis indicates that the baryon pole diagram always
leads to {\it an antibaryon tending to emerge parallel to the
outgoing meson in $\bar B\to \B_1\bar \B_2 M$ decays}
\cite{Chengreview,CCHT}. This feature has been confirmed in
$\Lambda_c^+\bar p\pi^-$ \cite{Belle:Lamcppi}, $p\bar p\pi^-$
\cite{Tetiana}\footnote{A recent study of angular distributions in
$B^-\to p\bar p\pi^-$ \cite{Geng06} indicates a correlation of the
pion with the proton. This is opposite to what we expect. This has
to be checked by future experiments at BaBar and Belle.}
and $\Lambda\bar p\gamma$ \cite{Belle:Lampgam} modes. However, the
measured angular distribution in $B^-\to p\bar pK^-$ turns out to
be astonishing.

Based on the pole model and the intuitive argument, the $K^-$ in
the $p\bar p$ rest frame is expected to emerge parallel to $\bar
p$. However, the Belle observation is other around
\cite{Belle:3charmless1}: the $K^-$ is preferred to move
collinearly with the proton in the $p\bar p$ rest frame. BaBar
\cite{BaBar:ppK} has studied the Dalitz plot asymmetry in the
invariant masses $m_{pK}$ and $m_{\bar pK}$ and found a result
consistent with Belle. This puzzle could indicate that the $p\bar
p$ system is produced from some intermediate states, such as the
glueball and the baryonium, a $p\bar p$ bound state, whcih may
change the correlation pattern. This possibility is currently
under study in \cite{CCHT}. (For a different treatment of the
correlation puzzle in $B^-\to p\bar p K^-$ decay, see
\cite{Geng06}.)

The three-body doubly charmed baryonic decay
$B\to\Lambda_c\bar\Lambda_cK$ has been observed recently by Belle
with the branching ratio of order $7\times 10^{-4}$
\cite{Belle:2LamcK}. Since this mode is color-suppressed and its
phase space is highly suppressed, the naive estimate of $\B\sim
10^{-8}$ is too small by four to five orders of magnitude compared
to experiment. It was originally conjectured in \cite{CCT} that
the great suppression for the $\Lambda_c^+\bar\Lambda_c^-K$
production can be alleviated provided that there exists a narrow
hidden charm bound state with a mass near the
$\Lambda_c\bar\Lambda_c$ threshold. This possibility is plausible,
recalling that many new charmonium-like resonances with masses
around 4 GeV starting with $X(3872)$  and so far ending with
$Y(4260)$ have been recently observed by BaBar and Belle. This new
state that couples strongly to the charmed baryon pair can be
searched for in $B$ decays and in $p\bar p$ collisions by studying
the mass spectrum of $D^{(*)}\ov D^{(*)}$ or
$\Lambda_c\bar\Lambda_c$. However, no new resonance with a mass
near the $\Lambda_c\bar\Lambda_c$ threshold was found by Belle
(see Fig. 3 in version 2 of \cite{Belle:2LamcK}). This implies the
failure of naive factorization for this decay mode and may hint at
the importance of nonfactorizable contributions such as
final-state effects. For example, the weak decay $B\to D^{(*)}\bar
D_s^{(*)}$ followed by the rescattering $D^{(*)}\bar D_s^{(*)}\to
\Lambda_c\bar\Lambda_c K$ \cite{CHChen} or the decay $B\to
\Xi_c\bar\Lambda_c$ followed by
$\Xi_c\bar\Lambda_c\to\Lambda_c\bar\Lambda_cK$ may explain the
large rate observed for $B\to\Lambda_c\bar\Lambda_cK$.

\section{Radiative baryonic $B$ decays}
Naively it appears that the bremsstrahlung process will lead to
$\Gamma(B\to\B_1\ov \B_2\gamma)\sim {\cal O}(\alpha_{\rm
em})\Gamma(B\to\B_1\ov \B_2)$ with $\alpha_{\rm em}$ being an
electromagnetic fine-structure constant and hence the radiative
baryonic $B$ decay is further suppressed than the two-body
counterpart, making its observation very difficult at the present
level of sensitivity for $B$ factories. However, there is an
important short-distance electromagnetic penguin transition $b\to
s \gamma$. Owing to the large top quark mass, the amplitude of
$b\to s\gamma$ is neither quark mixing nor loop suppressed.
Moreover, it is largely enhanced by QCD corrections. As a
consequence, the short-distance contribution due to the
electromagnetic penguin diagram dominates over the bremsstrahlung.

\begin{figure}[t]
\vspace{-1cm}
\hspace{0cm}\centerline{\includegraphics[width=19pc]{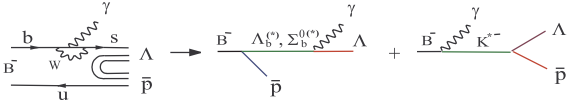}}
\vspace{-1.5cm}
    \caption{{\small Quark and pole diagrams for $B^-\to\Lambda\bar
    p\gamma$.
    }}
    \label{fig:Lampgammadia}
\end{figure}

Since a direct evaluation of this radiative decay is difficult as
it involves an unknown 3-body matrix element
$M_{\mu\nu}=\la\Lambda\bar p|\bar
s\sigma_{\mu\nu}(1+\gamma_5)b|B^-\ra$, we shall instead evaluate
the corresponding diagrams known as pole diagrams at the hadron
level (see Fig. \ref{fig:Lampgammadia}).
For
$B^-\to\Lambda\bar p\gamma$, the relevant intermediate states are
$\Lambda_b^{(*)}$, $\Sigma_b^{0(*)}$ and $K^*$.

The predicted branching ratios for $B^-\to\Sigma^0\bar p\gamma,~
\Xi^0\bar\Sigma^-\gamma$ and $\Xi^-\bar\Lambda\gamma$  decays are
summarized in Table \ref{tab:radBR} \cite{CYrad06}.
For the decay rates of other modes, see \cite{CYrad}. It is
interesting to notice that the $\Sigma^0\bar p\gamma$ mode, which
was previously argued to be very suppressed due to the smallness
of the strong coupling $g_{\Sigma_b\to B^-p}$ \cite{CYrad},
receives the dominant contribution from the $K^*$ pole diagram and
its branching ratio is consistent with that obtained in
\cite{GHrad}. In contrast, the mode $\Xi^0\bar\Sigma^-\gamma$ is
dominated by the baryon pole contribution. Meson and baryon
intermediate state contributions are comparable in $\Lambda\bar
p\gamma$ and $\Xi^-\bar\Lambda\gamma$ modes except that they
interfere constructively in the former but destructively in the
latter. Recently, Belle \cite{Belle:Lampgam} has made the first
observation of radiative hyperonic $B$ decay $B^-\to\Lambda\bar
p\gamma$ with the result
 \be
 \B(B^-\to\Lambda\bar p\gamma)=(2.16^{+0.58}_{-0.53}\pm0.20)\times
 10^{-6}.
 \en
In addition to the first observation of $\Lambda\bar p\gamma$, the
decay $B^-\to \Xi^0\bar\Sigma^-\gamma$ at the level of $6\times
10^{-7}$ may be accessible to $B$ factories in the future.

\begin{table*}[t]
\caption{Branching ratios and angular asymmetries for radiative
baryonic $B$ decays \cite{CYrad06}.} \label{tab:radBR}
\begin{center}
\begin{tabular}{| l l l l c |} \hline
 Mode & Baryon pole & Meson pole & Br(total) & Angular asymmetry \\ \hline
$B^-\to\Lambda\bar p\gamma$ & $7.9\times 10^{-7}$ & $9.5\times
10^{-7}$ & $2.6\times 10^{-6}$ & 0.25 \\
$B^-\to\Sigma^0\bar p\gamma$ & $4.6\times 10^{-9}$ &
$2.5\times 10^{-7}$ & $2.9\times 10^{-7}$ & 0.07 \\
$B^-\to\Xi^0\bar\Sigma^-\gamma$ & $7.5\times 10^{-7}$ & $1.6\times
10^{-7}$ & $5.6\times 10^{-7}$ & 0.43 \\
$B^-\to\Xi^-\bar\Lambda\gamma$ & $1.6\times 10^{-7}$ & $2.4\times
10^{-7}$ & $2.2\times 10^{-7}$ & 0.13 \\
\hline
\end{tabular}
\end{center}
\end{table*}

Besides the threshold enhancement effect observed in the
differential branching fraction of $\Lambda \bar p\gamma$, Belle
has also measured the angular distribution of the antiproton in
the $\Lambda\bar p$ system and found that the $\Lambda$ tends to
emerge opposite the direction of the photon. The angular asymmetry
is measured by Belle to be $A=0.36^{+0.23}_{-0.20}$ for
$B^-\to\Lambda\bar p\gamma$ \cite{Belle:Lampgam}. The meson pole
diagram is responsible for low-mass enhancement and does not show
a preference for the correlation between the baryon pair and the
photon. Our prediction $A=0.25$ (see Table \ref{tab:radBR}) is
consistent with experiment.

{\bf\noindent Acknowledgement}: I'm grateful to the organizers for
this wonderful conference.

\end{document}